# Experimental density radiography of Wudalianchi volcano with cosmic ray muons

Y. Cheng[a], R. Han[a], Z.Li[a], J.Li[a], J.Li[b], W.Gu[b], X.Yang[a], X.Ouyang[b], B.Liao[c]

[a] *Science and Technology on Reliability and Environmental Engineering Laboratory, Beijing Institute of Spacecraft Environment Engineering, Beijing 100094, China*

[b] *School of Nuclear Science and Engineering, North China Electric Power University, 2 BeiNong Road, Beijing 102206, China*

[c] *College of nuclear science and technology, Beijing Normal University, 19 Xinjiekou outer St., Beijing 100875, China*

E-mail: hanran@ncepu.edu.cn

S U M M A R Y

Muon radiography is a promising technique to image the internal density structures upto a few hundred meters scale, such as tunnels, pyramids and volcanos, by measuring the flux attenuation of cosmic ray muons after trvaling through these targets. In this study, we conducted an experimantal cosmic ray muon radiography of the Wudalianchi volcano in northeast China for imaging its internal density structures. The muon detector used in this study is made of plastic scintillator and silicon photomultiplier. After about one and a half month observation for the Laoheishan volcano cone in the Wudalianchi volcano, from September 23rd to November 10th, 2019, more than 3 million muon tracks passing the data selection criteria are obtained. Based on the muon observations and the high-resoluiton topography from aerial photogrammetry by unmanned aerial vehicle, the relative density image of the Laoheishan volcano cone is obtained. The experiment in this study is the first muon radiography of volcano performed in China, and the results suggest the feasibility of radiography technique based on plastic scintillator muon detector. As a new passive geophysical imaging method, cosmic ray muon radiography could become a promising method to obtain the high-resoution 2-D and 3-D density structures for shallow geological targets.

# 1. Introduction

Volcanic hazard assessment and risk management remain two very important scientific subjects with heavy implications both on the population safety and economic development. Anticipating future activity of volcanoes requires monitoring of their activity as well as the information on their internal structures.

In this contribution, we would like to use a new method which is known as muon radiography to investigate volcano inner structures. The basic idea of muon radiography is that muons will lose energy via ionization, radiation etc when they pass through mediums. Low energy muons will be terminated in the medium if all their kinetic energy is lost. The survival muon rate can be used to infer the average density in the pathway of the muon track. This proposal was brought up about 50 years ago (Luis 1970, Zhou 1987 ). As a results of the technology improvements in muon trackers, especially the plastic scintillator development, muon radiography is frequently applied in area of architecture, tunnel exploration, volcano imaging (Daniele 2014, Ambrosino, 2015, Morishima, 2017, Kunihiro, 2017, Ran 2020, ELENA, 2017, Hiroyuki 2009).

We carry out the first volcano muon radiography in Wudalianchi area in China. The Wudalianchi Volcano Field (WDF) is situated in Northeast China. It's about 1800 km away from the Pacific plate. The two neighbouring volcanoes is Changbaishan and Jingbohu volcanoes, 400 and 600 km away respectively. The most recent eruption of the WDF is around 300 years ago at the Laoheishan and Huoshaoshan volcano cones (Li，2016). The absolute height of Laoheishan is ∼ 166 m; the volcanic vent is about 350 m while the depth is 140 m. We would like to investigate the inner structure of Laoheishan volcano by muon radiography. A density map reflecting the average density in the muon pathway will be demonstrated.

**2. Method**

2.1 Muon Radiography Method

Muons, which are produced through the interaction of primary cosmic rays with earth atmosphere, has drawn a great deal of attention as an imaging method, called muon radiography. As muons passing through the medium, they will lose energy via their interaction with the medium. By doing the energy loss calibration, the density pathlength will be derived. By measuring the flux attenuation of muons as they pass through these targets, this can be used as an effective technique to image the density of the inner structure of a kilometer-scale geological object, such as tunnels, pyramids and volcanos (Nishiyama, 2014, Barnoud, 2019, Cosburn, 2019, Lelièvre, 2019).

2.2 Muon Detector

The muon telescope is a three-layer plastic scintillator detector with dimensions of 80cm × 5cm × 1cm. Each layer consists of two planes containing 16 plastic scintillator strips. The two sub-planes in each layer is placed orthogonally, to provide x and y coordinates for the muon fired spot. Details and readouts of the detector can be found in (Han, 2020).

2.3 Volcano Topography

Before the measurement, we did a geological survey by drone ( **Error! Reference source not found.**). A high-precision 3-D image of the volcano is derived, the horizontal resolution of the drone image is ~1 cm, and the vertical resolution is 10 cm. From the 3D image of the volcano, we calculated the path-length of the muon trajectory based on a stand-alone code. The path length was obtained from the detector point of view, see Figure 6.

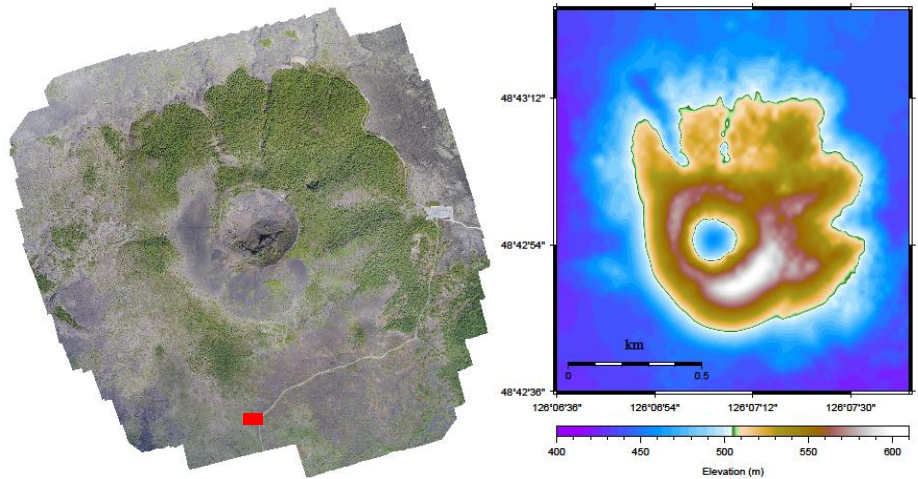

Figure 1 Left: Drone scan map of the Laoheishan region. The detector is placed close to the sightseeing road illustrated as a red square. Right: Elevation map of the Laoheishan volcano. We can see the relative height seen at the detector sight is ~200m.

Also, an elevation image is derived. The elevation of the detector is 400 m. The highest part of the volcano is ~ 600 m. So the relative height of the volcano is ~200 m, not a very huge volcano.

Based on the elevation image, we can get the density length image given in the geological coordinate frame whose horizontal axis denotes the azimuth angle and the vertical axis is the elevation angle, see the right plot of Figure 1. The path lengths are represented by different colours. This image presents the full range (elevation and azimuth) path length of the volcano, which provides us a precise estimation and baseline for thickness imaging of the volcano. Furthermore, according to this image, the detector is lifted up 20 degrees to do better scanning of the volcano. In this way, the acceptance of the detector is maximized, therefore, the data-taking time can be shortened. There are three stages in the whole data-taking procedure. First the orientation of the detector is towards northwest with 10 degrees to do volcano imaging. And then we rotate the detector to face the open sky where there is no volcano as baseline calibration run. The third rotation is after a quick look at the available data, and we find part of the image of the volcano is missing, therefore, we turn the detector to again to northeast 30 degrees to get a full image of the volcano.

## 3. Data analysis

### 3.1 Data Selection

Approximately 3 million muon tracks were acquired at Laoheishan site. Data taking was performed in time period of about one day, storing all the buffer events. In our measurement, the following selection method is used to determine the muon tracks among all the buffer events.

Supposing that the coordinates are (x1, y1), (x2, y2), (x3, y3) for plane 1, plane 2 and plane 3, these event lines can be determined by standard fitting procedure with the line cut,

$$|(x_1 - x_2) - (x_2 - x_3)| < 1$$
$$|(y_1 - y_2) - (y_2 - y_3)| < 1$$

where 1 denotes the id difference of the scintillation bars. Here we use very strict selection cuts to estimate the solid angle carefully.

To get the coordinates ($\theta_g, \phi_g$) in the geological coordinate frame of the muon track, Euler rotation was performed,

$$R_x = \begin{bmatrix} 1 & 0 & 0 \\ 0 & \cos\theta & -\sin\theta \\ 0 & \sin\theta & \cos\theta \end{bmatrix}$$

$$R_y = \begin{bmatrix} \cos\theta & 0 & \sin\theta \\ 0 & 1 & 0 \\ -\sin\theta & 0 & \cos\theta \end{bmatrix}$$

$$R_z = \begin{bmatrix} \cos\theta & -\sin\theta & 0 \\ \sin\theta & \cos\theta & 0 \\ 0 & 0 & 1 \end{bmatrix}$$

where $\theta$ denotes the corresponding rotation angle. For example, if the orientation of the detector is 20 degrees in the elevation and -10 degrees in the azimuth angle under geological coordinate frame, we first rotate 20 degrees with reference to y-axis and then rotate 10 degrees with reference to the z-axis to transform from the detector frame to the geological coordinate frame. In this way, we get the muon track direction in the geological coordinates.

**3.2 Open-sky data analysis**

To better understand the data, a simply Geant4 simulation package was developed. the differential muon flux is defined by the modified Gaisser's Formula (Guan, 2015). Compared to Gaisser's Formula which is applied to the high energy range (100 GeV -100 TeV (Lesparre:2010)) and small zenith angle, modified Gaisser's formula given in Eq. 1 corrects the low energy and large zenith angle flux distributions. The modified Gaisser's formula is given by

$$\Phi(\theta, E) = 0.14 \left(\frac{E^*}{GeV}\right)^{-2.7} \left(\frac{1}{1 + \frac{1.1E\cos\theta^*}{115 GeV}} + \frac{0.054}{1 + \frac{1.1E\cos\theta^*}{850 GeV}}\right) \quad (1)$$

Where

$$E^* = E\left(1 + \frac{3.64 GeV}{E(\cos\theta^*)^{1.29}}\right)$$

$$\cos\theta^* = \left(\frac{\cos^2\theta + p_1^2 + p_2(\cos\theta)^{p_3} + p_4\cos\theta^{p_5}}{1 + P_1^2 + P_2 + p_4}\right)$$

with $p_1 = 0.102573$, $p_2 = 0.068287$, $p_3 = 0.958633$, $p_4 = 0.0407253$, $p_5 = 0.817285$.

The predicted number of muons is as follows:

$$N_{in}(\theta) = \int_{0.1}^{\infty} \Phi(\theta, E) \cdot S_{eff}(\theta) \cdot \varepsilon \cdot \Delta T \, dE \quad (2)$$

$S_{eff}(\theta)$ denotes the effective area of the detector, $\varepsilon$ is the total efficiency in the experiment, $\Delta T$ is the data taking time, $\Phi(\theta, E)$ is the differential muon flux as a function of the zenith angle $\theta$ and the muon energy $E$ as shown above. The integration starts from 0.1 GeV, which is taken from experience.

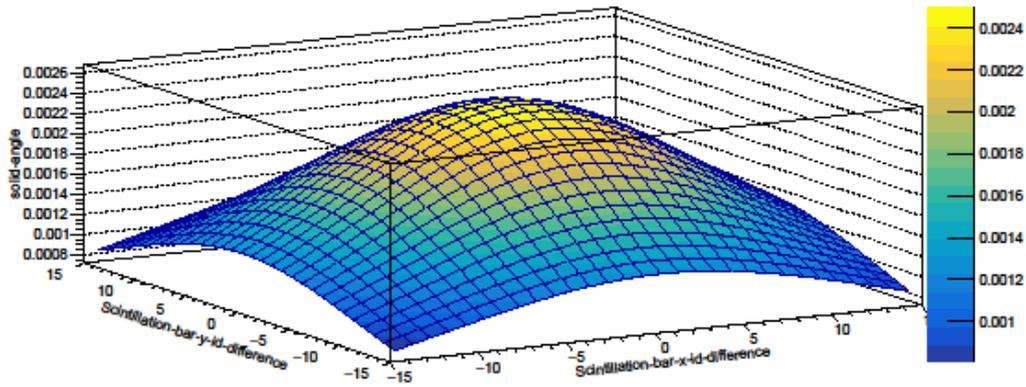

Figure 2 Solid angle as a function of the x-direction and y-direction id difference. The solid angle is used in the detection efficiency correction for muons with different incoming directions.

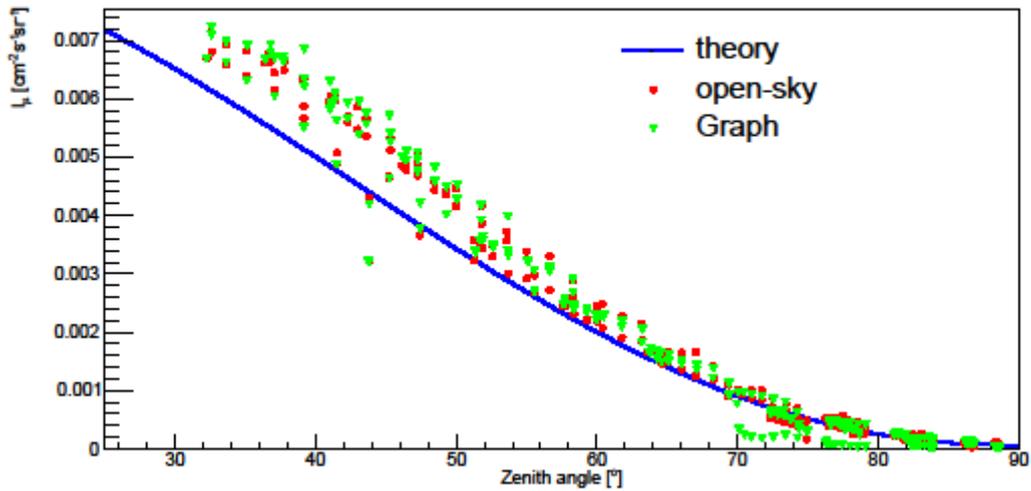

Figure 3 Muon flux measurement results in case of open-sky (red dot) and volcano (green triangle), and their comparison with the modified Gaisser formula prediction. The low zenith angle discrepancy is due to noise contamination, such as electrons, protons. In the high zenith angle range, we can clearly see the muon flux decrease due to the attenuation caused by the pass length in the volcano.

We did a muon flux dependence measurement with reference to the zenith angle, Figure 3. Various corrections were introduced taking into account the actual detector performance. These include the efficiency corrections of the liquid scintillation bars, the zenith angle dependent solid angle (Figure 2) and acceptance correction, also the detection area and time duration correction. Deviations from the modified Gaisser model can be seen in the lower zenith angle part in Figure 3, it's assumed to be caused by fake muon tracks which are indeed electrons or protons. To avoid this problem, it's proper to do some shielding, for example, using lead plates, to prevent some low energy protons or electrons from penetrating three layers of liquid scintillator bars. The exact flux number of the protons or electrons is hard to estimate, since our detector has no ability to do particle identification.

### 3.3 Volcano Data analysis

If there is a volcano, we need to modify the expected muon number as:

$$N_{in}(\theta) = \int_{E_{min}}^{\infty} \Phi(\theta, E) \cdot S_{eff}(\theta) \cdot \varepsilon \cdot \Delta T dE \quad (3)$$

where $E_{min}$ reflects the minimal kinetic energy that a muon must possess to hit the detector surface without being absorbed, which depends on the average medium density $\rho$ and the length muon passed through $l$. The transmission power of the muon is taken from (Groom, 1999). In this reference, the CSDA (continuously slowing down ability) for various materials can be found. Here we use liquid water as a reference to account for how muons lost its energy via ionization collision or radiation when passing-by liquid water.

For data analysis, muon transmission $K^a$ is defined as the ratio of the muon event rate recorded by the detector in the tunnel and that in the open air under the investigation, which can be calculated by

$$K^a = \frac{N_V(\theta)/\Delta T_V}{N_O(\theta)/\Delta T_O} \quad (4)$$

where subscripts $V, O$ denote data selected from the volcano direction and the open-sky, respectively. Substituting Eq. (1) into Eq. (4), we have

$$K^a = \frac{N_V(\theta)/\Delta T_V}{N_O(\theta)/\Delta T_O} = \frac{\int_{E_{min}}^{\infty} \Phi(\theta, E) dE}{\int_{E_0}^{\infty} \Phi(\theta, E) dE} \quad (5)$$

since $S_{eff}(\theta)$ and $\varepsilon$ do not change during the experiment.

By using Eq. (5) and muon energy spectrum described in modified Gaisser's formula as in Eq. (1). we can easily obtain $E_{min}$. The relationship between the minimum energy $E_{min}$ that muons must possess to penetrate the volcano without being absorbed and its density-length $R(E_{min})$ have been calculated by Groom, et al. (Groom, 1999), where the density of the standard rock is $\rho = 2.65 g/cm^3$ and $\frac{Z}{A} = 0.50$. In this paper, the density of the volcano rock is taken as $1.7 g/cm^3$. The density-length $R(E_{min})$ is defined by $R(E_{min}) = \rho \cdot L$. Above all, it is possible to get the functional relationship between $R(E_{min})$ and muon transmission $K^a$ power, ie the muon survival ratio. The calculated results are as follows and can be repeatedly used in future analysis.

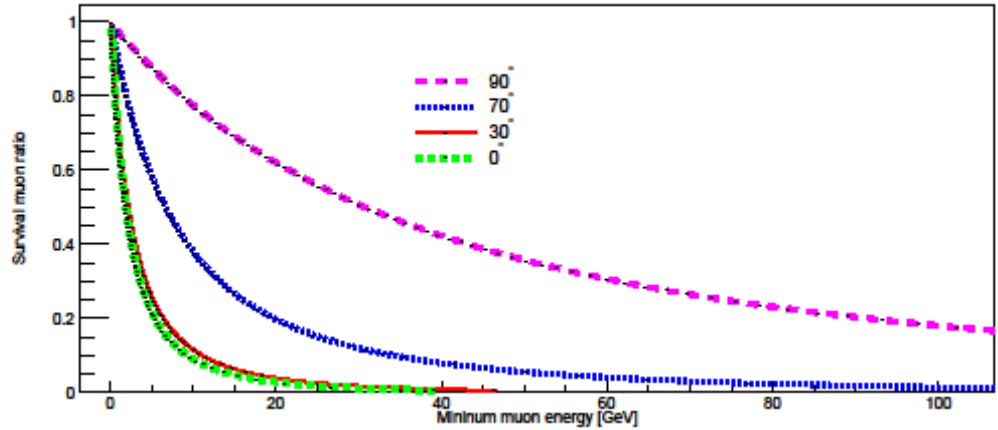

Figure 4 The survival muon ratio (transmission $K^a$) and the minimum muon energy under different zenith angle. The calculation is based on modified Gaisser formula. We can see that, the energy spectrum for horizontal muons is much harder than the vertical muons.

To get the mountain profile via muon radiography, we can directly use the flux ratio between the volcano and open-sky case. The ratio plot is obtained by the ratio of the elevation and azimuth angle dependent flux measurement results when the detector facing the volcano and facing the open-sky. The decrease in muon ratio reflects the average density of the volcano in the specific direction of the muon passage. For example, if the ratio equals 1.0, that means the density is the same, i.e. muons go merely through air in this specific direction. The raw data-taking time normalized muon flux ratio is shown as in top and bottom plot. In this plot, only data-taking time difference is corrected. The detector related correction, such as efficiency correction of scintillation bars, solid angle and acceptance correction can be cancelled out naturally in the case of pixel by pixel muon flux ratio calculation. In Figure 5, we can clearly see the volcano profile represented by the red part. Since the whole data-taking period is composed of three stages, one is northwest 10 degrees, towards the west part of the volcano. The other is open-sky when detector faces no volcano, and the third one is facing northeast 30 degree, towards the east part of the volcano.

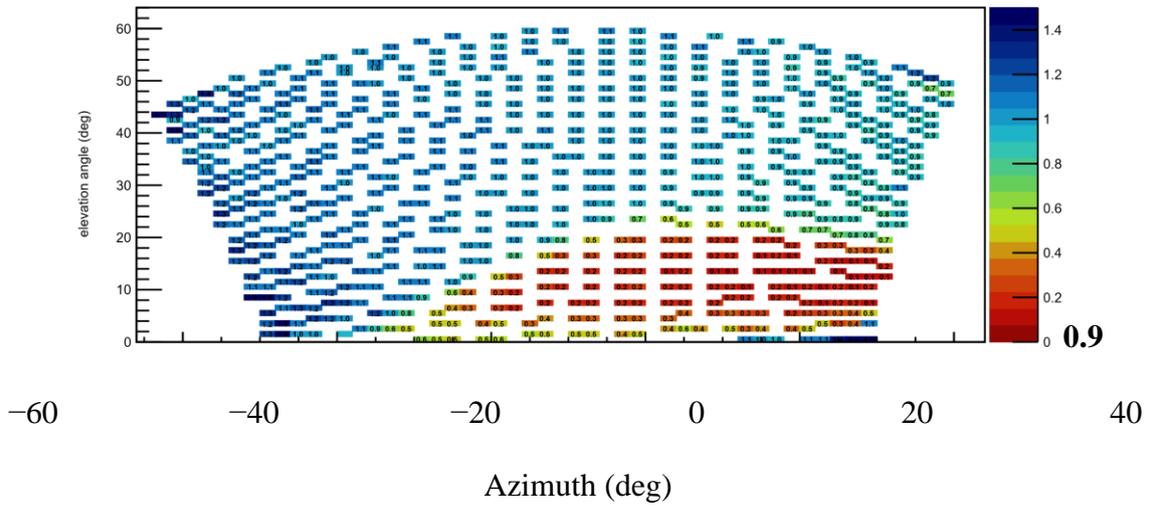

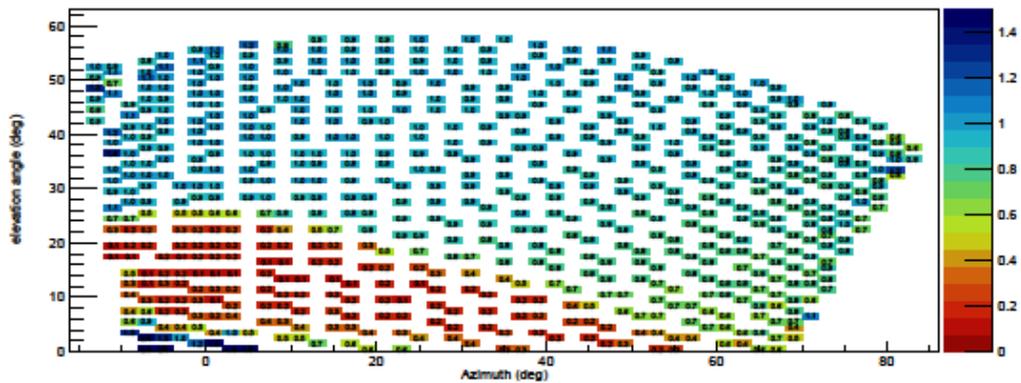

Figure 5 Raw data-taking time normalized ratio plot from the data. The top one is the ratio when detector facing northwest 10 degrees and open-sky, and the west part of the volcano is clearly represented by the small ratio. The bottom plot is the ratio when detector facing northeast 30 degrees and open-sky, similarly the east part of the volcano is clearly represented by the decreased ratio. The blue part in both the top and bottom plot, where the ratio is close to 1, means there is no decrease in muon rate, which means there is no volcano in the passage of the muons in this specific direction.

From the ratio plot, we can see the impact from the volcano, due to the density length of the volcano, some of the muon tracks with energy smaller than minimum required energy is terminated inside the volcano, thus we can see ratio less than 1. For the specific elevation and azimuth angle, where there is no volcano, we can see the ratio value is very close to 1.0 with small uncertainty.

However, we can see that the ratio is much larger than the expectation value (taking into account the drone scan results, where the volcano is hundreds of meters in length, and assume the average density of the volcano is ~ 1.7 $g/cm^3$). Discussions are made in the following. First of all, the noise from other radioactive sources act as backgrounds, which can also pass through all the three layers of the plastic scintillators. Second, in the analysis, we treat the detector as a point. Then it's possible that parallel muons in certain direction may or may not pass through the volcano. The parallel muons which don't pass through the volcano also acts as background. However, calculation indicates that parallel muons effects can hardly affect the ratio. Thirdly, considering the geometrical of the detector, the backward muons can also increase the ratio, for horizontal muons, we cannot differentiate which side the muons actually come from, that's the possible reason why in low elevation angle where the path length in volcano is the largest, such as 0 degree, the ratio increase again. Last but not least, the low energy muons which are scattered at the surface of the volcano, and the changed their direction can also be taken as the high energy muons passing through volcano but with a wrong direction, therefore gives us a wrong ratio value.

However, when overlay the ratio plot with the path-length inside the volcano plot, we can see good boundary agreement between these two plots. The path length inside the volcano and the ratio is in good coincidence taking into account the relatively large size of the scintillation bars, i.e. ~5 cm, and the various background. The separation between air and the volcano is obvious, with blue part represents the air and the colourful part represents the volcano.

After the ratio data analysis, we obtained the profile imaging of Laoheishan volcano, see **Error! Reference source not found.**. However, we are more interested in the inner structure of the volcano. Density abnormal is one of the key issues we would like to study.

The procedure of density unfolding is as follows. First, we have the ratio value with azimuth and elevation angle set. From the modified Gaisser formula, we can calculated the threshold of the muon energy, serving as the minimum energy to pass through the volcano, under this zenith angle. Then according to the stopping power curve from (Groom, 1999), we can infer the path length inside the certain material. In this manuscript, we use the CSDA curve for the liquid water (1 g/cm$^3$), and the derived density length if using CSDA curve from standard rock (2.65 g/cm$^3$) is very similar. Since we already have the path length from the drone scanning data, we can deduce the density directly by dividing the path length.

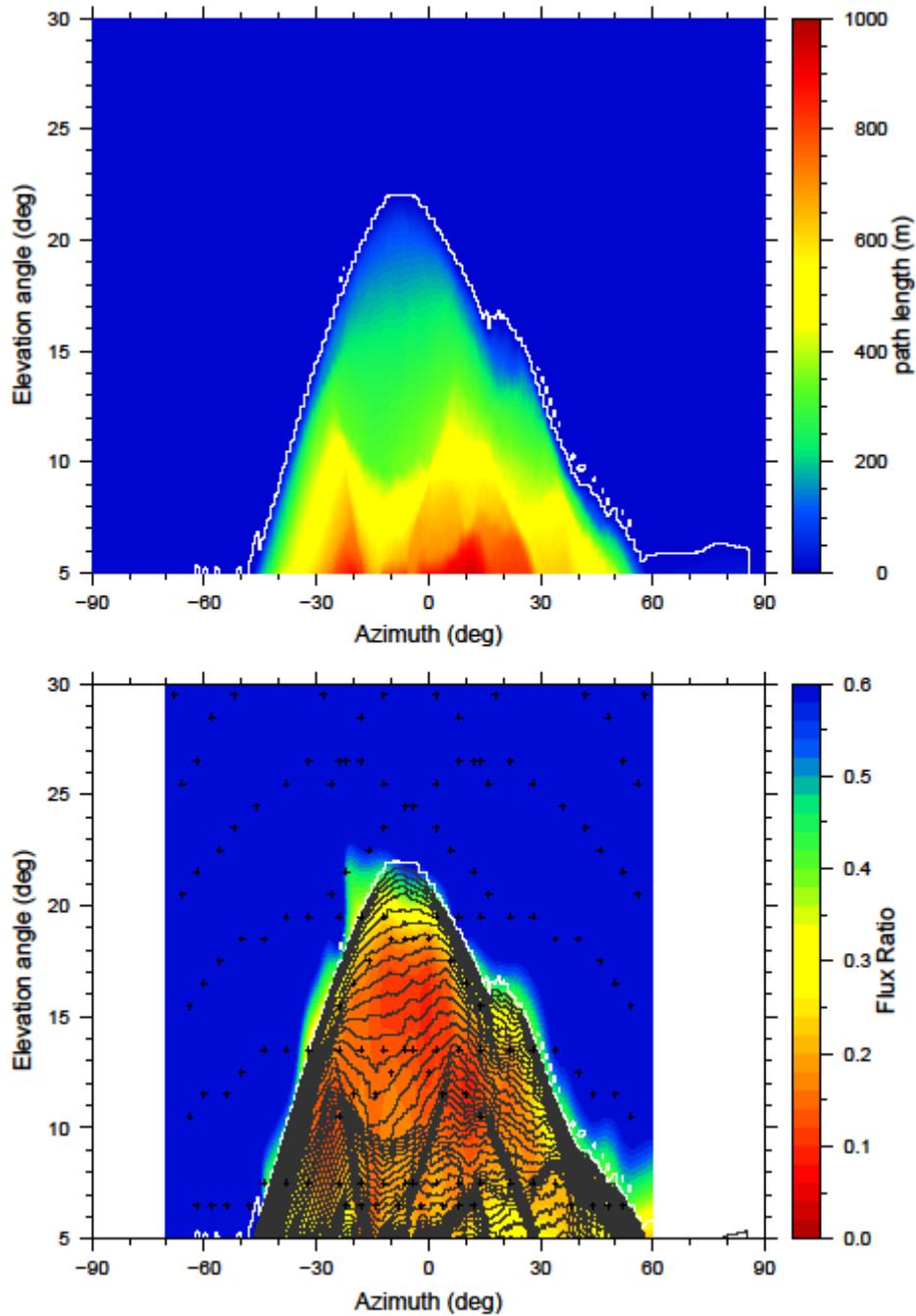

Figure 6 Top: the path length inside the volcano calculated from the drone scan data. The white line clearly visible in the plot is the 1meter equal length contour. The maximum path length is close to 1000 meters, which almost stops all the muons. Bottom: the overlaid ratio plot and the path length plot. We can see clear boundary separation between air and volcano.

**4. Results and Discussions**

Following the procedures introduced in the previous section, we can get the density map. However, as a result of the too high ratio value, the unfolded density is too low and has no physics meaning. Here we publish the relative density results, ignoring the exact numbers. The results are shown as Figure 7.

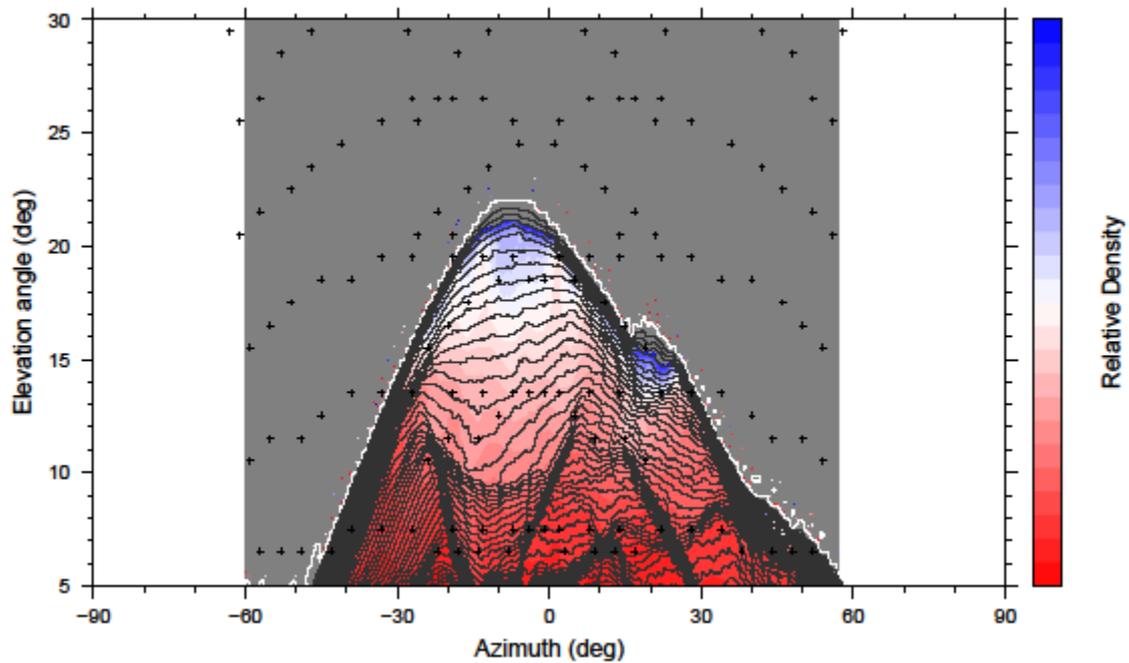

Figure 7 The unfolded density plot of the volcano. Due to noise, the ratio value is not correct. And as a result, the unfolded density result is too low to have physical meaning. Here we only publish the relative density. In the plot, we can see some density pattern is presented: the top of the volcano has relatively large density and the bottom of the volcano

has relatively small density. This is similar to the observation by (Tanaka,2007). They also observed high density region beneath the dome.

Since the measured ratio between volcano scenario and open-sky scenario is one order larger than expected, we can only have the relative density map inside the volcano. We raised up the following factors that can influence the measurement. The electronic noise from the SiPM or readout board. The electrons or protons that pass through all the 3 layers and form a straight line, and behaves as a muon track. There are also some backward muons, which comes from the side without volcano, especially the incoming direction of the horizontal muons can not be distinguished. The scattered muons is another very important background, it's muons that passing through the edge of the volcano and scattered by the volcano, and changed its original direction. As a result, the scattered muons smear the angular distribution of the muons after passing through the volcano.

Another reason which lead to the unphysical density resutls is that the violent temperature change during the data-taking period. During the data-taking period, the temperature in Laoheishan region goes from 20 Celsius degree to minus 10 Celsius degree. The daily event rate fluctuation is within 10% range at three different experiment stage. SiPM with automatic temperature compensation function is more desirable.

To suppress the fake muon tracks, such as electrons and protons, lead plate between scintillator layers is suggested. The lead plate can terminate some low energy electrons or protons. Besides, it can reflect the direction of high energy electrons and protons. Check the direction before and after the lead plate can be used as a method to do electron and muon discrimination.

For the back-ward muons, considering the practical condition of the observation field, muons from backward relative to the volcano direction can dilute the effective signal as mentioned before. To evaluate the backward muon noises, we used a Genat4 simulation package to estimate the noise level. The simulation results suggest that muons from frontward versus muons from backward is 100:1.494 in the whole zenith angle range, and contributed hugely in large zenith angle range.

The scattered muons can be very annoying and ruined the measurement. (Nishiyama, 2016) has put up a method to do evaluate scattered muon noise with some Geant 4 simulation.

Due to the backgrounds and noise mentioned above, using muon radiography in actual volcano internal density measurement can be still challenging and should be considered in R&D process of the muon detector.

**5. Conclusion**

In this study, we conducted the first muon radiography of volcano in China. In case of volcano, we can clearly see muon flux defect in the zenith angle range covered by the volcano, ie 70-90. We did the relative ratio measurement and successfully get the profile image of the volcano. The boundary of the volcano is in well coincidence with the drone scaned image. The seperation of air and volcano is clear.

We also had the relative density to inspect the inner structures of the volcano. The relative density map reflects that the top of the volcano is high-density region and the bottom of the volcano is low density region. (Tanaka, 2007) has similar observation.

In summary, the volcano works as a filter to modulate the muon flux and angular distribution. It requires good background control and evalution, and can be used as a method to derive the inner structure of the kilo-meter scale object.

**Acknowledgements**

This work was supported by the National Laboratory Foundation (Grant No. 6142004180203).